\documentclass[amsmath,amssymb,aps,prd, preprintnumbers,nofootinbib,a4paper, 11pt]{revtex4-1}
\usepackage[utf8]{inputenc}
\usepackage{hyperref}
\hypersetup{
    colorlinks=true,
    linkcolor=black,
    filecolor=blue,      
    urlcolor=blue,
    citecolor=blue
}
\usepackage{xcolor}
\usepackage{youngtab}
\usepackage{amsmath}
\usepackage{ytableau}
\usepackage{ amssymb }

\begin{document}

\title{{ \bf \huge
Spectrum of anomalous dimensions in hypercubic theories}}

\author{Oleg Antipin,}
\email{oantipin@irb.hr} 
\author{Jahmall  Bersini}
\email{jbersini@irb.hr} 
\affiliation{Rudjer Boskovic Institute, Division of Theoretical Physics, Bijeni\v cka 54, 10000 Zagreb, Croatia}

\begin{abstract}
We compute the spectrum of anomalous dimensions of non-derivative composite operators with an arbitrary number of fields $n$ in the $O(N)$ vector model with cubic anisotropy at the one-loop order in the $\epsilon$-expansion. The complete closed-form expression for the anomalous dimensions of the operators which do not undergo mixing effects is derived and the structure of the general solution to the mixing problem is outlined. As examples, the full explicit solution for operators with up to $n=6$ fields is presented and a sample of the OPE coefficients is calculated. The main features of the spectrum are described, including an interesting pattern pointing to the deeper structure.
\end{abstract}

\maketitle
\tableofcontents

\section {Introduction}

Conformal field theories (CFTs) play a fundamental role in our understanding of the Universe with applications ranging from the discovery of the asymptotic freedom in QCD \cite{Gross:1973id,Politzer:1973fx} to the potential applications in gravity in the scenario of asymptotic safety \cite{weinberg}. Mathematically, for a given symmetry group, CFT is specified via \textit{CFT data}, i.e. the scaling dimensions of all the primary operators and the set of OPE coefficients, defined as the constants appearing in the three point functions of the theory. 

In this work, we study the CFT data of the theory invariant under the hypercubic symmetry group $H_N$ realized as a group of symmetries of an $N$-dimensional 
hypercube. In nature, this symmetry group appears in the description of critical properties of cubic magnets, like iron~\cite{Pelissetto:2000ek}, in which the magnetic anisotropy induced by the lattice structure is experimentally accessible. 
It also appears in the description of certain structural phase transitions such as the cubic to tetragonal transition in $SrTiO_3$ (\textit{strontium titanate})\cite{Aharony}. 

Hypercubic models are usually investigated in the perturbative $\epsilon$-expansion 
\cite{Wilson} 
using standard diagrammatic techniques and present-day results involve the computation of anomalous dimensions ($\gamma_\phi$, $\gamma_{m^2}$) and beta functions to six loops order \cite{Carmona:1999rm,Adzhemyan:2019gvv}.
Recently this theory was also explored non-perturbatively via an exact renormalization-group equation \cite{Vidal} or resorting to the  the conformal bootstrap method \cite{Stergiou:2018gjj, Kousvos:2018rhl, Rong:2017cow,Dey:2016mcs}. Most of these studies have focused on $H_3$ \cite{Stergiou:2018gjj, Kousvos:2018rhl, Rong:2017cow}\footnote{Interestingly, these non-perturbative results are in tension with those obtained from the $\epsilon$-expansion.}, with the exception of \cite{Dey:2016mcs}, which computed CFT data for composite operator with two fields and arbitrary spin at $\mathcal{O}(\epsilon^3)$ for a generic $N$. Finally, the model was also used in \cite{Fei:2015oha} to investigate the generalized $F$-theorem conjecture.

The aim of this paper is to compute the spectrum of anomalous dimensions for composite operators with an \emph{arbitrary number of fields $n$} but no derivatives in the $H_N$ critical theory to $\mathcal{O}(\epsilon)$. 
Our results are complementary to previous works which usually consider small values of $n$ (and/or fix $N$) but at the same time consider arbitrary number of derivatives and higher order in the $\epsilon$-expansion. After introducing the model in Sec.\ref{model} we lay out
the representation theory for $H_N$ in Sec.\ref{Irreducible representations of $H_N$} with the aim to summarize it in a practical way useful for future investigations of hypercubic models. 
For the computation of the anomalous dimensions we make use of a recently developed techniques \cite{Rychkov:2015naa, Codello:2018nbe} based on the Schwinger–Dyson equation and CFT constraints, which we outline in Sec.\ref{Computation}. 
In Sec.\ref{towers} we present our main results, providing a
complete formula for anomalous dimensions of an infinite number of composite operators with arbitrary $n$ which do not mix with the others and supply an algorithm to obtain the solution to the remaining mixing problem. In Sec.\ref{Spectrum of anomalous dimension} we show explicit results for the anomalous dimension of composite operators up to $n = 6$ and we give a qualitative description of the spectrum for $n > 6$.
Sec.\ref{cftdata} discusses the way to obtain the rest of the CFT data as a further step towards completing the knowledge of the theory and include the computation of some OPE coefficients as an example. We give our conclusions in Sec.\ref{conclusion}.

\section{$O(N)$ model with cubic anisotropy} \label{model}

The Lagrangian for the $O(N)$ Ginzburg-Landau model with cubic anisotropy in $d = 4 - \epsilon$ is: 
\begin{equation}
    \mathcal{S}_{GL}=\int D^{d}x \left(\frac{(\partial \phi_i)^2}{2}+\frac{g_1}{4!}(\phi_i\phi_i)^2+\frac{g_2}{4!} \sum_{i} \phi_i^4 \right)
\end{equation}
where the interaction terms can be rewritten in a tensor form as
$ \tfrac{1}{4!}   V_{ijkl} \phi_i \phi_j \phi_k \phi_l$
with
\begin{equation} \label{eq4}
    V_{ijkl}= \frac{g_1}{3}(\delta_{ij}\delta_{kl}+\delta_{il}\delta_{kj}+\delta_{ik}\delta_{jl})+g_2 \delta_{ijkl}
\end{equation}
and the tensor $\delta_{ijkl}$ defined by
\begin{equation}
    \delta_{ijkl}=
    \begin{cases}
    1, \ \ \text{when} \ \ i=j=k=l \\ 
    0, \ \ \text{otherwise} \ .
    \end{cases}
\end{equation}
The presence of the coupling $g_2$ breaks the $O(N)$ symmetry so that the action is invariant only under $H_N \subset O(N)$ where $H_N$ is the group of symmetries of an $N$-dimensional hypercube.  

The renormalized model predicts four fixed points (FPs) which, at the 1-loop level, read
\begin{eqnarray}
  && (g_1^G,g_2^G)=\left(0 , 0 \right), \qquad \qquad \qquad \quad \ \ (g_1^I,g_2^I)=(4 \pi)^2\left(0,\frac{\epsilon}{3}\right),   \nonumber \\
  && (g_1^O,g_2^O)=(4 \pi)^2\left(\frac{3\ \epsilon}{N+8},0 \right), \qquad (g_1^H,g_2^H)=(4 \pi)^2\left(\frac{\epsilon}{N},\frac{(N-4)}{3N} \epsilon\right) .
\end{eqnarray}
The first is a trivial Gaussian FP while the second and the third correspond, respectively, to the $\phi^4$ theory ($N$ decoupled Ising models) and to the O($N$) model. The fourth, often called "cubic fixed point", corresponds to a theory invariant under $H_N$, and the main goal of this paper is to compute the spectrum of anomalous dimensions at this last FP.

For this purpose, it is important to note that for $N = N_c = 4$ the cubic FP coincides with the $O(N)$ symmetric one, for $N \rightarrow \infty$ it correspond to the Ising one, while for $N=1$ there is only one coupling constant $g_1 + g_2$ and the cubic FP reduces to the free theory. As we will see later, these limits will provide non-trivial cross-checks for our results. 

A further consistency check is given by the existence of a special symmetry for $N=2$. In fact the interaction term is invariant if we perform a $\frac{\pi}{4}$ rotation of the fields~\cite{H. Kleinert and V. Schulte-Frohlinde},
\begin{equation}
    \phi_1^{\prime}=\frac{\phi_1+\phi_2}{\sqrt{2}}, \ \ \ \
    \phi_2^{\prime}=\frac{\phi_1-\phi_2}{\sqrt{2}}
\end{equation}
and at the same time transform the coupling constants with
\begin{equation}
    g_1^{\prime}=g_1+\frac{3}{2}g_2, \ \ \ \ g_2^{\prime}=-g_2 \ .
\end{equation}
This turns the cubic FP into the Ising one. 

While the values of $N$ which relate the cubic FP to the Gaussian and Ising ones do not depend on the order of calculation in perturbation theory, the value of $N_c$, for which the cubic FP coincides with the $O(N)$ symmetric one, does depend on it. A precise determination of $N_c$ is physically of great interest because for $N < N_c$ the infrared physics is controlled by the $O(N)$ FP while for $N > N_c$ the cubic critical regime is realized. 
The most recent results \cite{Adzhemyan:2019gvv} from the 6-loop calculation suggest $N_c \sim 2.9$, implying that the cubic FP could be relevant for describing the physical $N=3$ case of cubic ferromagnets.

For our purpose, it is also useful to note that, as a consequence of the equations of motion (EOM) at the cubic fixed point:
\begin{equation}\label{eq1}
    \Box \phi_i = \frac{1}{{3 !}} \left(g_1 \phi_i \phi^2 +g_2 \phi_i^3  \right), \qquad \phi^2 \equiv \sum_{i} \phi_i^2
\end{equation}
the operator $g_1 \phi_i \phi^2 +g_2 \phi_i^3$\ , which is a \textit{primary} operator in the free theory, becomes a  \textit{descendant} (a derivative) of the primary operator $\phi_i$ in the interacting theory, and thus its properties are entirely fixed in terms of those of $\phi_i$.
This phenomenon is called \textit{conformal multiplet recombination} \cite{Rychkov:2015naa} and in particular it implies that:
\begin{equation} \label{eq8}
    \Delta_{(  \phi_i \phi^2 +\frac{g_2}{g_1} \phi_i^3 )}=\Delta_{\phi_i} + 2 \quad \Longrightarrow \quad  \gamma_{  (\phi_i \phi^2 +\frac{g_2}{g_1}\phi_i^3) }=1 \ ,
\end{equation}
where the scaling dimensions of a composite operator with $n$ fields $S_n$ is given by:
\begin{equation}
    \Delta_{S_n}=n(1-\frac{\epsilon}{2})+\gamma_{S_n} \epsilon + \mathcal{O}(\epsilon^2)
\end{equation}

with $\gamma_{S_n}$ the 1-loop anomalous dimension. We will use \eqref{eq8} as an additional check for our results.

Having introduced the model under consideration, the next step is to unveil the operator spectrum at the cubic fixed point, i.e. to find the irreducible representations of the hypercubic group. 

\section{Irreducible representations of hypercubic group $H_N$} \label{Irreducible representations of $H_N$}

The aim of this section is to lay out the representation theory for the hypercubic group in a way suited 
for a streamlined construction of the spectrum of $H_N$ composite operators as we are not aware of such construction and few relevant results are somewhat scattered in the literature \cite{Baake,Hamermesh,JamesKerber1981}.
%

$H_N$ is a discrete subgroup of $O(N)$ and is given by the wreath product $\mathcal{S}_N \ltimes \mathcal{Z}_2^N$. To construct the irreducible representations of $\mathcal{S}_N \ltimes \mathcal{Z}_2^N$, we start by computing the outer tensor products of the irreducible representations of $\mathcal{Z}_2$ $N$ times. Labeling as $[1^2]$ and $[2]$ the two irreducible representations of $\mathcal{Z}_2$, the irreducible representations of $\mathcal{Z}_2^N$ are
\begin{equation} \label{Z2N}
    [2]^{\otimes \alpha}\otimes[1^2]^{\otimes \beta} \ , \qquad \alpha+\beta=N \ .
\end{equation}
In accordance with these representations, the symmetric group $S_N$ is divided into direct product $\mathcal{S}_\alpha \times \mathcal{S}_\beta$ and then the irreducible representations of $\mathcal{S}_N \ltimes \mathcal{Z}_2^N$ are generated by multiplying those of $\mathcal{Z}_2^N$ in Eq.\eqref{Z2N} with the corresponding direct product \footnote{In the language of group theory, the irreducible representation of $\mathcal{Z}_2^N$ have to be multiplied with those of the corresponding \textit{inertia factor groups}.} $\mathcal{S}_\alpha \times \mathcal{S}_\beta$ \cite{Balasubramanian}.
For instance, the irreducible representations of $\mathcal{Z}_2^3$ are:
\begin{equation}
        [2]^{\otimes 3}, \ [2]^{\otimes 2}\otimes[1^2], \ [2]\otimes[1^2]^{\otimes 2}, \ [1^2]^{\otimes 3}
\end{equation}
and they have to be multiplied by the irreducible representations of  $\mathcal{S}_3$, $\mathcal{S}_2 \times \mathcal{S}_1$, $\mathcal{S}_1 \times \mathcal{S}_2$ and $\mathcal{S}_3$, respectively.
From this construction, it is clear that an elegant way to label the irreducible representations of $H_N$ is in terms of double-partitions of N, $(\alpha,\beta)$, which can be represented as ordered pairs of Young diagrams with $\alpha$ and $\beta$ boxes, respectively \cite{orellana, JamesKerber1981}. For example, the ten irreducible representations of $H_3$ are
\begin{eqnarray}
  & (\bf [2]^{\otimes 3} \otimes \mathcal{S}_3) : ( {\scriptsize \Yvcentermath1  \yng(3)}  ,{\bf \emptyset}  )  ,  ( {\scriptsize \Yvcentermath1  \yng(2,1)}  ,{\bf \emptyset}  ) ,  ( {\scriptsize \Yvcentermath1  \yng(1,1,1)}  ,{\bf \emptyset}  ) ;  \quad (\bf [2]^{\otimes 2}\otimes[1^2] \otimes \mathcal{S}_2 \times S_1) : ( {\scriptsize \Yvcentermath1  \yng(2)}  ,{\scriptsize \Yvcentermath1  \yng(1)}  )  , ( {\scriptsize \Yvcentermath1  \yng(1,1)}  ,{\scriptsize \Yvcentermath1  \yng(1)}  )  ;\nonumber \\
  & ( \bf [1^2]^{\otimes 3} \otimes \mathcal{S}_3): (  {\bf \emptyset}  , {\scriptsize \Yvcentermath1  \yng(3)}  )  , (  {\bf \emptyset}  , {\scriptsize \Yvcentermath1  \yng(2,1)}  )  , (  {\bf \emptyset}  , {\scriptsize \Yvcentermath1  \yng(1,1,1)}  ) ; \quad (\bf [2] \otimes[1^2]^{\otimes 2} \otimes \mathcal{S}_1 \times S_2) : ( {\scriptsize \Yvcentermath1  \yng(1)}  ,{\scriptsize \Yvcentermath1  \yng(2)}  )  , ( {\scriptsize \Yvcentermath1  \yng(1)}  ,{\scriptsize \Yvcentermath1  \yng(1,1)}  )  \nonumber
\end{eqnarray}
where "$\emptyset$" stands for the empty partition. The left partition represents $\alpha$ objects, \textit{even} under $\mathcal{Z}_2$, which transform under an irreducible representation of $\mathcal{S}_{\alpha}$, while the right partition represents $\beta$ objects, now \textit{odd} under $\mathcal{Z}_2$, which transform under an irreducible representation of $\mathcal{S}_{\beta}$. 

The dimension of a given double-partition $(\alpha,\beta)$ is \cite{orellana}
\begin{equation} \label{eq6}
  dim(\alpha,\beta)=\left( {\begin{array}{cc}
  N  \\
  \alpha \\
  \end{array} } \right) \times dim(\alpha) \times dim(\beta)
\end{equation}
where $dim(\alpha)$ ($dim(\beta)$) is the  dimension of the corresponding representation of the  symmetric group $S_\alpha$ ($S_\beta$) obtained via the standard hook's rule.

The defining $N-$dimensional representation of $H_N$ is given by
\begin{equation}
  \phi_i=(N-1,1)=( \ \underbrace{{\small \Yvcentermath1  \yng(2)}\dots \small \Yvcentermath1\yng(1)}_{N-1} \ , \ {\small \Yvcentermath1  \young(i)} \ ) 
\end{equation}
and the decomposition of the tensor product of the defining representation with an arbitrary representation $(\alpha,\beta)$ can be obtained through the formula:
\begin{equation}
  (N-1,1) \otimes (\alpha,\beta) = \sum_{\alpha^+,\ \beta^-} (\alpha^+,\beta^-) \oplus \sum_{\alpha^-,\ \beta^+} (\alpha^-,\beta^+)
  \label{generaltensor}
\end{equation}

where $\alpha^+$ are the tableaux obtained by moving one box from $\beta$ to $\alpha$ and $\alpha^-$ are the tableaux obtained moving one box from $\alpha$ to $\beta$. The same procedure applies to $\beta^+$ and to $\beta^-$. For example, for the tensor product $\phi_i \otimes \phi_j$, we have
\begin{equation} \label{secondlevel}
    (  \underbrace{{\scriptsize \Yvcentermath1  \yng(2)}...}_{N-1} \ ,  {\scriptsize \Yvcentermath1  \yng(1)}  )   \otimes  (  \underbrace{{\scriptsize \Yvcentermath1  \yng(2)}...}_{N-1}  \ ,  {\scriptsize \Yvcentermath1  \yng(1)}  )  =  (  \underbrace{{\scriptsize \Yvcentermath1  \yng(2)}...}_{N} \ ,  {\bf \emptyset}  ) \oplus 
    (  \underbrace{{\scriptsize \Yvcentermath1  \yng(2,1)}...}_{N} \ , {\bf \emptyset}  ) \oplus (  \underbrace{{\scriptsize \Yvcentermath1  \yng(2)}...}_{N-2}  \ ,  {\scriptsize \Yvcentermath1  \yng(2)}  ) \oplus (  \underbrace{{\scriptsize \Yvcentermath1  \yng(2)}...}_{N-2} \ ,  {\scriptsize \Yvcentermath1  \yng(1,1)} )
\end{equation}

Now, we are in position to construct explicitly the composite operator of $H_N$ with $n$ fields and no derivatives. The first step is to find the corresponding bi-tableaux by computing the tensor product of the defining representation $n$ times following the rules of \eqref{generaltensor}. 
As a result, we will have some bi-tableaux that never appeared before at smaller $n$ and each such bi-tableau will correspond to a \textit{unique} composite operator. In addition, we will also have bi-tableaux that already appeared at the levels $n-2$, $n-4$, ..., and "reappeared" at the level $n$ simply as a consequence of moving one box from one side of double-partition to the other and then returning it back to the original place.
In this case, more than one composite operator of order $n$ will correspond to that tableau, and to find the true scaling operators, it will be necessary to solve the mixing problem.

The second step is to associate to every bi-tableau the corresponding $H_N$ tensor as follows.
Let us assume, for the moment, that we are dealing with unique composite operator and consider some generic representation having $0$ boxes in the right partition and $N$ boxes in the left partition ($\alpha=N,\ \beta=0\ $). For example:
\ytableausetup
  {mathmode, boxsize=.85em,aligntableaux=center}
 $ \left( \ \begin{ytableau} 
\empty  &\empty  &\none[...]&  \scriptstyle   \\
&  \\
\empty & \none[]  \\
 \end{ytableau}  {\huge{,}} \
 \emptyset
\  \right) $  \ .\\

We start by filling the left tableau with indices that we impose to be all different: 
\begin{center}
\ytableausetup
  {mathmode, boxsize=1em, aligntableaux=center}
\qquad \qquad \qquad $\left(  \ \begin{ytableau}
\scriptstyle \mu_1 &\scriptstyle \mu_2 &\none[...]& \scriptstyle \scriptstyle \mu_{s} \\
i& j\\
k & \none[]  \\
 \end{ytableau}  {\huge{,}} 
 \ \emptyset
\  \right)   \qquad  i \neq j \neq k \neq  \mu_1 \neq \mu_2  \neq...\neq \mu_{s} \ . $ \\
  \end{center} 
Since the left partition represents the objects even under $Z_2$, then for every box in the first row we associate an indexed field raised to the \textit{zeroth} power; for every box in the second row, we associate an indexed field raised to the \textit{second} power and so on, in order of increasing \textit{even} powers of the fields for subsequent rows. Finally, we have to symmetrize all these indices as usual (symmetrization over all boxes in the same row, and antisymmetrization over all boxes in the same column). This leads to ($n=8$ in this case)
\begin{eqnarray}
&\ytableausetup
  {boxsize=1em, aligntableaux=center}
  \left( \  \begin{ytableau}
\scriptstyle \mu_1 &\scriptstyle \mu_2 &\none[...]& \scriptstyle \scriptstyle \mu_{s} \\
i& j\\
k & \none[]  \\
 \end{ytableau}  {\huge{,}} 
 \ \emptyset
 \  \right)   
\quad = \quad (\phi_{[k}^4\phi_i^2\phi_{\mu_1]}^0)\cdot (\phi_{[j}^2\phi_{\mu_2]}^0)\cdot \phi_{\mu_3}^0 \phi_{\mu_4}^0...\phi_{\mu_{s}}^0 = \nonumber \\
   & = (\phi_k^4 \phi_i^2+\phi_{\mu_1}^4 \phi_k^2+\phi_i^4 \phi_{\mu_1}^2-\phi_k^4 \phi_{\mu_1}^2-\phi_i^4 \phi_k^2-\phi_{\mu_1}^4 \phi_i^2)\ (\phi_j^2 - \phi_{\mu_2}^2) \  \nonumber
 & \quad i \neq j \neq k \neq \mu_1 \neq \mu_2  \nonumber
\end{eqnarray}
The same rules also apply to the right partition $\beta$, but now we have to associate increasing \textit{odd} powers of the fields as we increase the number of rows. For instance (with $n=5$) 
\begin{eqnarray*}
&&\left(\young({{\mu_1}}{{\mu_2}}{{\mu_3}})\dots \young({{\mu_s}}) \ , \ {\small \Yvcentermath1  \young({i}{k},{j})} \ \right) 
= \underbrace{\phi_{\mu_1}^0 \phi_{\mu_2}^0...\phi_{\mu_{s}}^0}_{\text{left}} \times \underbrace{\phi_k (\phi_i^3\phi_j-\phi_j^3\phi_i)}_{\text{right}}= \nonumber\\
   && = \phi_k (\phi_i^3\phi_j-\phi_j^3\phi_i) \ \ \ \ \  i \neq j \neq k \ . \nonumber
\end{eqnarray*}
Certain bi-tableaux can also appear for the first time at a level $n$ too low to allow the previous constructions and this simply means that the corresponding tensor requires derivatives to be built.\footnote{This is the case for the last irreducible representation in Eq.\eqref{secondlevel}, whose corresponding $H_N$ tensor cannot be built with only two fields.}
Aside from that, the most general $H_N$-tensor will be represented by the bi-tableau with the left and the right partitions separately comprising the most general Young diagram associated with the corresponding symmetric group.
It will have $k$ different types of columns (distinguished by the number of boxes) that we label with the index $i$. Each type of column can have multiplicity $p_i$. All the indices for the fields have to be different, and the indices of the fields in each column have to be antisymmetrized. The columns associated to the left partition will contain the even powers of the fields while the columns associated to the right partition will contain the odd ones and we label the highest power of the field in the given column by $m_i$.
Thus the unique hypercubic composite scaling operators corresponding to this most general bi-tableau can be written compactly as
\begin{eqnarray} \label{eq5}
  H_{n,\{m_i\}, \{p_i\}} =\prod_{i=1}^{k}   ( \phi_{[{\mu^i}_1}^{m_i} \phi_{{\mu^i}_2}^{m_i-2} \phi_{{\mu^i}_3}^{m_i-4}... \phi_{{\mu^i}_{q_i}]}^M)^{p_i}  \qquad \mu^i_1 \neq \mu^i_{2} \neq ... \neq \mu^i_{q_i}
\end{eqnarray}
where
\begin{equation}
    M=\begin{cases}
   0 \ \ \text{if} \ \ m \ \  \text{is} \ \ \text{even} \\  1 \ \  \text{if} \ \ m \ \ \text{is}  \ \ \text{odd} 
\end{cases} \ .
\end{equation}

Now, let us assume that we are dealing with a bi-tableau that "reappeared" at the level $n$. The corresponding mixing space can be found through the following steps:
\begin{enumerate}
    \item  
    Write the \emph{unique} composite operator corresponding to the bi-tableau according to the rules above, and then multiply the result with the appropriate power of $\phi^2$ needed to reach the level $n$. For instance, for $n=6$, we have:
\begin{equation*}
    ( \ \underbrace{{\small \Yvcentermath1  \yng(4,1)}...}_{N} \ , \ {\bf \emptyset} \ ) = (\phi^2)^2(\phi_i^2-\phi_j^2) \ \ \ \ \  i \neq j  \ . 
\end{equation*} 
  \item  Then "distribute" $\phi^2$ through the rest of the tensor in all possible ways as follows: 
\begin{equation*}
    ( \ \underbrace{{\small \Yvcentermath1  \yng(4,1)}...}_{N} \ , \ {\bf \emptyset} \ ) = \begin{cases}
    (\phi^2)^2(\phi_i^2-\phi_j^2) \\
    (\phi^2)(\phi_i^4-\phi_j^4) \\
    (\phi_i^6-\phi_j^6) \\
    \end{cases}  \ \ \ \ \  i \neq j  \ .
\end{equation*}
\item Finally, it is also necessary to take into account the mixing between powers of $\phi^2$ and the other $H_N$-scalars. The $H_N$-scalars of order $n$ are formed by products and powers of all the operators of the form
\begin{equation} \label{scalarmix}
    \sum_{i} \phi_i^2 = \phi^2, \ \sum_{i} \phi_i^4, \ \sum_{i} \phi_i^6,...,\sum_{i} \phi_i^n \ .
\end{equation}
For our example this means that $(\phi^2)^2$ will mix with $\sum_{i} \phi_i^4$ so that one additional operator has to be added to the mixing space:
\begin{equation*}
    ( \ \underbrace{{\small \Yvcentermath1  \yng(4,1)}...}_{N} \ , \ {\bf \emptyset} \ ) = \begin{cases}
    \sum_{k} \phi_k^4(\phi_i^2-\phi_j^2) \\
    (\phi^2)^2(\phi_i^2-\phi_j^2) \\
    (\phi^2)(\phi_i^4-\phi_j^4) \\
    (\phi_i^6-\phi_j^6) \\
    \end{cases}  \ \ \ \ \  i \neq j  
\end{equation*}
\end{enumerate}

In the next section we will show how to solve the mixing, find the true scaling operators of $H_N$ and compute their anomalous dimensions.

\section{Computation} \label{Computation}

To compute the spectrum of anomalous dimensions $\gamma$ we resort to a method first proposed in \cite{Rychkov:2015naa} (and recently generalized in \cite{Codello:2018nbe}) which makes use of constraints from conformal symmetry combined with the Schwinger-Dyson equation (SDE) in order to obtain non-trivial consistency conditions which allow us to extract the value of $\gamma$.
The key idea is to consider three-point functions of the form
\begin{equation} \label{eq2}
    \langle  \Box \phi_i S_n S_{n+1} \rangle
\end{equation}
and use the EOM \eqref{eq1} to rewrite them as \footnote{If we are interested only in the leading order in $\epsilon$ we can simply consider the classic EOM instead of the SDE.}

\begin{equation} \label{eq3}
   \langle  (g_1\phi_i \phi^2 +g_2 \phi_i^3 ) S_n S_{n+1} \rangle
\end{equation}
Here, $S_n$ is a composite operator of order $n$, i.e. a product of $n$ fields transforming under an irreducible representation of $H_N$. 

Matching the results for the computation of the three-point functions Eqs.\eqref{eq2} and \eqref{eq3}, gives a recursion relation for the anomalous dimension of the scaling operators $\gamma_{S_n}$, whose general solution at the cubic fixed point is the following eigenvalue equation:
\begin{equation} \label{eq9}
    \gamma_{S_n} S_{i_1,i_2,..,i_n}=\frac{n(n-1)}{32 \pi^2}V_{j_1,j_2,(i_1,i_2} S_{i_3,..,i_n),j_1,j_2} \ .
\end{equation}

Here, $V_{j_1,j_2,i_1,i_2}$ is the tensor defined in Eq.\eqref{eq4} and the tensors $S$ are given by:
\begin{equation}
\label{Stensor}
   S_{i_1,i_2,..,i_n}=\tfrac{1}{n!} \partial_{i_1} \partial_{i_2}...\partial_{i_n}S_n
\end{equation}
Equation \eqref{eq9} can be recast in a more practical form
\begin{equation} \label{eigenvalue}
\mathcal{D}S_n =\gamma_{S_n}S_n
\end{equation}
where \footnote{Equation \eqref{eigenvalue} with $\mathcal{D}$ given by Eq.\eqref{eq7} was already found in \cite{Osborn:2017ucf}.} :
\begin{equation} \label{eq7}
\mathcal{D}=\frac{1}{3N}\left(\frac{ \phi^2 \partial^2 }{2}+(\phi \cdot  \partial)^2 - \phi \cdot  \partial + \frac{N-4}{2} \sum_i \phi_i^2 \partial_i^2 \right)
\end{equation}

The scaling operators are found with the techniques described in the previous section. For the operators which mix Eq.\eqref{eigenvalue} determines uniquely the mixing matrix.

\section{Towers} \label{towers}

In order to cross-check our results for anomalous dimensions of $H_N$ operators, we need to recall the corresponding ones for $O(N)$ and Ising models.
\subsection{$O(N)$ }

At the $1$-loop order, for $N = N_c =4$ the cubic FP coincides with the $O(N)$ symmetric one. The $O(N)$ operators with no derivatives belong to the $O(N)$ fully symmetric space, which at the level $n$ is
\begin{equation}
\label{symmON}
  \underbrace{ {\small \Yvcentermath1  \yng(4)}...\small \Yvcentermath1  \yng(1)}_{n}+ \underbrace{ {\small \Yvcentermath1  \yng(3)}...{\small \Yvcentermath1  \yng(1)}}_{n-2}+\underbrace{ {\small \Yvcentermath1  \yng(2)}...\small \Yvcentermath1  \yng(1)}_{n-4}+...+ \begin{cases}  {\bf Tr } \ \ \text{if} \ \ n \ \  \text{is} \ \ \text{even} \\  {\small \Yvcentermath1  \yng(1)} \ \ \ \text{if} \ \ n \ \ \text{is}  \ \ \text{odd}  \end{cases}
\end{equation}
and the dimension of this space is given by the binomial coefficient:
\begin{equation}
\label{fulldimON}
  \left( {\begin{array}{cc}
  N+n-1  \\
  n  \\
  \end{array} } \right) \ .
\end{equation}
Some of the results in Sec.\ref{Irreducible representations of $H_N$} can be deduced from the representation theory for $O(N)$. For instance, it is clear from Eq.\eqref{symmON} that any $O(N)$ tableau which appears for the first time at a level $n$ will "reappear" at the levels $n+2$, $n+4$, ..., and the fact that we observed this happening also in $H_N$ is a simple consequence of $H_N \subset O(N)$. Also, the fact that some $H_N$ tensors require derivatives to be explicitly constructed means that they do not come from the decomposition of operators belonging to the fully symmetric space of $O(N)$.

In the $O(N)$ model, the scaling operators with $n$ fields and no derivatives are m-index traceless symmetric tensors $T_{i_1...i_m}^{(m)} \phi^{2q}$, with $m=n-2q$, which correspond to the $m$-boxes $O(N)$ tableaux with one row. 
Application of the CFT techniques described in Sec.\ref{Computation} allows the computation of $1$-loop anomalous dimensions for the entire "tower" of $O(N)$ scaling operators \cite{Gliozzi:2017hni,Osborn:2009vs,Kehrein:1994ff},

 





\begin{equation}
\label{towerON} 
    \gamma_{m,q}=\frac{m(m-1)+q(N+6(q+m)-4)}{N+8} \ 
\end{equation}
with the corresponding dimension given by

\begin{equation}
\label{dimON}
dim= \left( {\begin{array}{cc}
   N+m-1  \\
   m  \\
  \end{array} } \right)-\left( {\begin{array}{cc}
   N+m-3  \\
   m-2  \\
  \end{array} } \right) \ .
\end{equation}

During our analysis, we found it useful to check the decomposition of $O(N)$ tensors to those of $H_N$ by monitoring the dimension of the representations using Eqs.\eqref{eq6} and \eqref{dimON}. Explicit examples of this decomposition will be presented in Sec.\ref{Spectrum of anomalous dimension}.

\subsection{Decoupled Ising model}
For $N = 2,\infty$ the cubic fixed point coincides with $N$ decoupled copies of the Ising model. In this case only the representations coming from the smallest Young tableaux in Eq.\eqref{symmON} survive which is a ${\bf Tr }$ if $n$ is even and $\scriptsize \yng(1)$ if $n$ is odd. Since in the decoupled Ising model the field has only one component, these representations altogether simply collapse to the "tower" of  composite scaling operators $\phi^n$, and their leading order anomalous dimensions in $d=4-\epsilon$, are given by \cite{Rychkov:2015naa}
\begin{equation}
\label{towerising} 
    \gamma_{n}=\frac{n(n-1)}{6} \ . 
\end{equation}

\subsection{The hypercubic tower}

Finally, we turn our attention to the anomalous dimensions at the cubic fixed point. As discussed above, at every $n$, there will be a set of bi-tableaux which appear for the first time and each such bi-tableau corresponds to only one scaling operator. This class of operators is given by Eq.\eqref{eq5} and they will compose our "hypercubic tower".
The corresponding 1-loop anomalous dimensions can be found by applying the differential operator $\mathcal{D}$ defined in Eq.\eqref{eq7},
\begin{equation}
    \mathcal{D}H_{n,\{m_i\}, \{p_i\}} =\gamma_{n,\{m_i\}, \{p_i\} }H_{n,\{m_i\}, \{p_i\}}
\end{equation}
and for the most general bi-tableau having $k$ different types of columns (labeled by the index $i$) with multiplicity $p_i$ and the highest power of the field in the column given by $m_i$, we obtain
\begin{equation} \label{gammas}
\boxed{\gamma_{n,\{m_i\}, \{p_i\} }= \frac{1}{6N} \left(2n(n-1)+(N-4) \sum\limits_{i=1}^{k} p_i [m_i(m_i-1)+(m_i-2)(m_i-3)+...]\right)}
\end{equation}
This is the main result of this paper. The dimensions of the operators in this tower are given by Eq.\eqref{eq6}. 
All the composite operators of order $n$ in Eq.\eqref{eq5} come from the decomposition of the longest ($n$-boxes) symmetric $O(N)$ tableau in Eq.\eqref{symmON}. In fact, for $N = 4$, Eq.\eqref{gammas} matches Eq.\eqref{towerON} with $q=0$, providing an explicit check.


For a first example of operators in Eq.\eqref{eq5}, we consider the simple case $k=1$, $m_i=1$, and $p_i=p=n$. This is given by the following bi-tableaux with one row
\begin{equation}
   ( \ \underbrace{{\small \Yvcentermath1  \yng(4)}...}_{N-p} \ ,\ \underbrace{{\small \Yvcentermath1  \yng(4)}...}_{p} \ )
\end{equation}
and corresponds to the following tower of operators:
\begin{equation}
  \phi_{\mu_1}  \phi_{\mu_2}...  \phi_{\mu_p} \qquad \mu_1 \neq \mu_2 \neq ... \neq \mu_p \ .
\end{equation}
The dimensions and eigenvalues of the operators in this tower at the level $n=p$ are
\begin{eqnarray}
&& dim=
  \left( {\begin{array}{cc}
   N  \\
   n  \\
  \end{array} } \right) \qquad \qquad
 \gamma_n= \frac{n(n-1)}{3N} \ .
\end{eqnarray}
For a less trivial example, we consider the family of bi-tableau with two rows, corresponding to $k=3$, $m_i=\{1,2,3 \}$ and $p_i=\{p_1,p_2,p_3 \}$, where the first and second rows of the left (\textit{right}) partition have respectively $N-p_1-p_2-p_3$ and $p_2$ (\textit{$p_1$ and $p_3$}) boxes.
This defines the tower of operators
\begin{align}
    & \underbrace{  \phi_{\mu_1} \phi_{\mu_2}... \phi_{\mu_{p_1}}}_{p_1 \ \text{terms}} \underbrace{( \phi_{\mu_{p_1+1}}^2-\phi_{\mu_{p_1+2}}^2)( \phi_{\mu_{p_1+3}}^2-\phi_{\mu_{p_1+4}}^2)...( \phi_{\mu_{p_1+2p_2-1}}^2-\phi_{\mu_{p_1+2p_2}}^2)}_{p_2 \ \text{terms}} \times   \nonumber \\ 
    & \times \underbrace{ ( \phi_{\mu_{p_1+2p_2+1}}^3 \phi_{\mu_{p_1+2p_2+2}}-\phi_{\mu_{p_1+2p_2+2}}^3 \phi_{\mu_{p_1+2p_2+1}})....( \phi_{\mu_{q-1}}^3 \phi_{\mu_q}-\phi_{\mu_q}^3 \phi_{\mu_{q-1}})}_{p_3 \ \text{terms}}, \quad \mu_1 \neq \mu_2 \neq ... \neq \mu_q
\end{align}
where
\begin{equation}
  q=p_1+2p_2+2p_3 \qquad \qquad n=p_1+2p_2+4p_3 \ .
\end{equation}
The dimensions and the $\gamma$'s of the operators in this tower at the level $n$ are
\begin{equation}
  dim=
  \left( {\begin{array}{cc}
  N \\
  p_2  \\
  \end{array} }\right) \left( {\begin{array}{cc}
  N-p_2 \\
  2p_3+p_1  \\
  \end{array} }\right) 
  \left( {\begin{array}{cc}
  2p_3+p_1\\
  p_3+p_1  \\
  \end{array} }\right)
  \frac{(N-p_1-2p_3-2p_2+1)(p_1+1)}{(N-p_1-2p_3-p_2+1)(p_1+p_3+1)} 
\end{equation}
and
\begin{equation} \label{torre2}
\gamma_{n,p_1,p_2}= \frac{4n(n-1)+(N-4)(3n-2p_2-3p_1)}{12N} \ .
\end{equation}
 
 Aside from Eq.\eqref{gammas}, it is clear that the content of the previous sections provides an algorithmic way to compute $\gamma$ for every non-derivative composite operator of arbitrary order $n$. However, as we will see,
 at large $n$, the mixing of operators corresponding to the same double-partition becomes complex, making the analytic progress difficult. 
 In the next section we will present explicit results up to $n=6$.
 
\section{Spectrum of anomalous dimensions} \label{Spectrum of anomalous dimension}

The indices of the operators in this section, with exception of those in Sec.\ref{n=1}, have to be understood to be all different, e.g. $\phi_i \phi_j$ means $\phi_i \phi_j$ together with the condition $i \neq j$.

\subsection{$n=1$} \label{n=1}

The anomalous dimension of $\phi_i$ appears only at $\mathcal{O}(\epsilon^2)$. In the spirit of Sec.\ref{Computation}, it can be computed at the leading order by considering the following two point function
\begin{equation}
  \Box_x \Box_y \langle \phi_i(x) \phi_j(y) \rangle
\end{equation}
and using the EOM \eqref{eq1} to rewrite it as
\begin{equation} \label{eqnn1}
   \langle  (g_1\phi_i \phi^2 +g_2 \phi_i^3 )(x)(g_1\phi_j \phi^2 +g_2 \phi_j^3 )(y)  \rangle
\end{equation}
Matching the evaluation of these two point functions gives
\begin{equation}
  \Delta_{\phi_i}=1-\frac{\epsilon}{2}+\frac{(N+2)(N-1)}{108N^2}\epsilon^2+\mathcal{O}(\epsilon^3) \ .
\end{equation}

\subsection{$n=2$}

The anomalous dimensions of the composite operators with two fields are

$\begin{array}{ccc}
  dim( \ \underbrace{ {\tiny \Yvcentermath1  \yng(4)}...}_{N-2}\ , \ {\tiny \Yvcentermath1  \yng(2)} \ )=\left( \scriptsize {\begin{array}{cc}
  N  \\
  2 \\
  \end{array} } \right)  & \qquad \phi_i \phi_j  &\qquad \gamma = \frac{2}{3N}  \nonumber\\ 
  dim( \  \underbrace{ {\tiny \Yvcentermath1  \yng(4,1)}...}_{N}\ , \ {\bf \emptyset}\ )= (N-1)  &\qquad  {\phi_i}^2- {\phi_j}^2  &\qquad \gamma = \frac{N-2}{3N} \nonumber\\
  dim( \  \underbrace{ {\tiny \Yvcentermath1  \yng(4)}...}_{N}\ , \ {\bf \emptyset}\ )=1  & \qquad 
  \phi^2  &\qquad \gamma = \frac{2(N-1)}{3N} 
\end{array} $

As expected, the dimensions of these representations add up to $\frac{N(N+1)}{2}$ which is Eq.\eqref{fulldimON} for $n=2$, while the 2-index traceless symmetric $O(N)$ tensor decomposes under $H_N$ as
\small \begin{equation}
    { \Yvcentermath1  \yng(2)}=\frac{(N+2)(N-1)}{2}={N-1}\oplus{\frac{N(N-1)}{2}} \ .
\end{equation} \normalsize
i.e. into a traceless diagonal symmetric tensor ${\phi_i}^2- {\phi_j}^2$ and an off-diagonal symmetric tensor $\phi_i \phi_j$. The results in this subsection were already obtained in \cite{Dey:2016mcs}.

\subsection{$n=3$} \label{n=3}

For the operators with $n=3$ fields, we obtain

$\begin{array}{ccc}
  dim( \ \underbrace{ {\tiny \Yvcentermath1  \yng(4)}...}_{N-3}\ , \ {\tiny \Yvcentermath1  \yng(3)} \ )= \left( \scriptsize{ {\begin{array}{cc}
  N  \\
  3 \\
  \end{array} } }\right)  &\qquad  \phi_i \phi_j \phi_k  &\qquad \gamma = \frac{2}{N}  \nonumber\\  
  dim(\  \underbrace{ {\tiny \Yvcentermath1  \yng(4,1)}...}_{N-1}\ , \ {\tiny \Yvcentermath1  \yng(1)} \ )= N(N-2)  &\qquad \phi_i( {\phi_j}^2- {\phi_k}^2 )  &\qquad \gamma = \frac{2+N}{3N}  \nonumber\\ 
  dim( \ \underbrace{ {\tiny \Yvcentermath1  \yng(4)}...}_{N-1}\ , \ {\tiny \Yvcentermath1  \yng(1)} \ )= N  &\qquad  
  \phi_i \phi^2 -2 \phi_i^3  &\qquad \gamma = \frac{2(N-1)}{3N}   \nonumber\\ 
  dim( \ \underbrace{ {\tiny \Yvcentermath1  \yng(4)}...}_{N-1}\ , \ {\tiny \Yvcentermath1  \yng(1)} \ )= N  &\qquad  
  \phi_i \phi^2 +\frac{N-4}{3} \phi_i^3  &\qquad \gamma = 1 
\end{array}$

The dimensions listed above add up to $\frac{N(N+1)(N+2)}{6}$, as they should from Eq.\eqref{fulldimON}. The 3-index traceless symmetric $O(N)$ tensor decomposes under $H_N$ as
\begin{equation}
  \small { \Yvcentermath1  \yng(3)}=\frac{N(N-1)(N+4)}{6}=N\oplus {N(N-2)}\oplus{\frac{N(N-1)(N-2)}{6}} \ .
\end{equation} \normalsize
To the best of our knowledge, the results of this subsection are new. They pass all the consistency checks as follows.
For $N=4$, they give the corresponding $O(N)$ results given by Eq.\eqref{towerON}: for the first three operators, we take $q=0$ and $m=3$ as they decompose the 3-index traceless symmetric $O(N)$ tensor, while for the last operator, $q=m=1$, as it is an $O(N)$ vector. For $N=1$, all the cubic operators vanish except the third one that gives the free theory result \footnote{Interestingly, the fourth operator listed does not vanish by its dimension but because the eigenvector itself becomes identically $0$.} $\gamma=0$. For $N = 2$ and $N=\infty$,
the cubic fixed point coincides with Ising one. In this case, only last operator remains, which reproduces the Ising model result for $\gamma$ in Eq.\eqref{towerising}. Finally, for this operator, Eq.\eqref{eq8}, which comes from the multiplet recombination phenomenon, is also satisfied.

\subsection{$n=4$}

The results for operators with four fields are

$\begin{array}{ccc}
  dim= \left( \scriptsize{ {\begin{array}{cc}
  N  \\
  4 \\
  \end{array} }} \right)\normalsize  & \ \ \phi_i \phi_j \phi_k \phi_l  & \qquad \gamma = \frac{4}{N}  \nonumber\\  
  dim= \frac{N(N-1)}{2}  & \ \ ( \phi_i \phi_j^3-\phi_j \phi_i^3 )  & \qquad \gamma = 1  \nonumber\\ 
  dim=\frac{N(N-3)}{2}  &  \ \
  (\phi_i^2 -\phi_j^2)(\phi_k^2 -\phi_l^2)  & \qquad \gamma = \frac{2(N+2)}{3N}   \nonumber\\
 dim= \frac{N(N-1)(N-3)}{2}  &  \ \
  \phi_i \phi_j (\phi_k^2-\phi_l^2)  & \qquad \gamma = \frac{N+8}{3N}  \nonumber\\ 
  dim= \frac{N(N-1)}{2}  & \ \ \frac{12-N+\sqrt{N^2+24N-48}}{12} 
  (\phi_i \phi_j \phi^2)+ \phi_i\phi_j^3+\phi_j \phi_i^3   & \qquad \gamma = \frac{12+5N + \sqrt{N^2+24N-48}}{6N}  \nonumber\\ 
dim= \frac{N(N-1)}{2}  & \ \
 \frac{12-N-\sqrt{N^2+24N-48}}{12}  (\phi_i \phi_j \phi^2)+ \phi_i\phi_j^3+\phi_j \phi_i^3   & \qquad \gamma = \frac{12+5N - \sqrt{N^2+24N-48}}{6N}  \nonumber\\ 
  dim= (N-1)  &  \ \
 \frac{4-N}{2} (\phi_i^2 -\phi_j^2 )\phi^2+\phi_i^4- \phi_j^4   & \qquad \gamma = 1 \nonumber\\ 
 dim= (N-1)  &  \ \
   \frac{4}{3}(\phi_i^2 -\phi_j^2 )\phi^2+\phi_i^4- \phi_j^4  & \qquad \gamma = \frac{2(3N-2)}{3N} \nonumber\\ 
  dim= 1  &  \ \
  (\phi^2)^2-2 \sum_{i=1}^N \phi_i^4  & \qquad \gamma = \frac{4(N-1)}{3N} \nonumber\\ 
    dim=1  & \ \
  (\phi^2)^2+\frac{N-4}{3} \sum_{i=1}^N \phi_i^4  & \qquad \gamma = 2  \\ \\
\end{array}$

The sum of the dimensions of these representations is $\frac{N(N+1)(N+2)(N+3)}{24}$ while the decomposition of the 4-index traceless symmetric $O(N)$ tensor under $H_N$ is
\small \begin{eqnarray}
  && { \Yvcentermath1  \yng(4)}=\frac{N(N+6)(N+1)(N-1)}{24}=1\oplus N-1\oplus 2\times \frac{N(N-1)}{2}\oplus\frac{N(N-3)}{2} \nonumber \\ 
  &&\oplus {\frac{N(N-1)(N-3)}{2}}\oplus{\frac{N(N-1)(N-2)(N-3)}{24}} \ .
\end{eqnarray} \normalsize
These results were already obtained in \cite{Osborn:2017ucf} and here we already solved the mixings explicitly. The reader may easily check the consistency of the results for $N=1,2,4,\infty$.

\subsection{$n=5$}

For the case of composite operators with $n=5$ fields, we have

$\begin{array}{ccc}
  dim= \left( \scriptsize { {\begin{array}{cc}
   N  \\
   5 \\
  \end{array} } } \right)  &  \phi_i \phi_j \phi_k \phi_l \phi_m   & \gamma = \frac{20}{3N}   \nonumber\\ 
   dim= \frac{N(N-1)(N-4)}{2}  & \phi_i (\phi_j^2 - \phi_k^2)( \phi_l^2- \phi_m^2) & \gamma = \frac{2(N+6)}{3N}  \nonumber\\ 
dim= \frac{N(N-1)(N-2)}{3}  &  
  \phi_i ( \phi_j^3\phi_k -\phi_k^3\phi_j) & \gamma = \frac{3N+8}{3N}  \nonumber\\ 
dim= \frac{N(N-1)(N-2)(N-4)}{6}  &  
  \phi_i \phi_j  \phi_k( \phi_l^2 -\phi_m^2) & \gamma = \frac{16+N}{3N}  \nonumber\\ 
dim= 2\times \frac{N(N-1)(N-2)}{6}  &  
  {\small   \left( {\begin{array}{cc}
 {\phi}^2 \phi_i \phi_j \phi_k  \\ 
  \phi_i^3 \phi_j \phi_k + perm  
  \end{array} } \right)} & \gamma = \frac{30+5N \pm \sqrt{(46+N)(N-2)}}{6N}  \nonumber\\
 dim=3 \times N(N-2)  &  
  {\small  \left( {\begin{array}{cc}
  {\phi}^2  \phi_i (\phi_j^2 - \phi_k^2)  \\ 
  \phi_i^3 (\phi_j^2 - \phi_k^2)  \\
  \phi_i (\phi_j^4 - \phi_k^4)  \\ 
  \end{array} } \right)} & \gamma = \frac{1}{3N}
  \scriptsize \left( {\begin{array}{ccc}
  3(N+6) & 3 & 6 \\
  2(N-4) & 4(N+1) & 0 \\ 
  4(N-4) & 0 & 2(3N-2)  
  \end{array} } \right) \normalsize     \nonumber\\ 
  \end{array}$
  
   $\begin{array}{ccc}
 dim=4 \times N  &  \qquad \qquad \qquad {\small \left( {\begin{array}{cc}
  \phi_i \left({\phi}^2 \right)^2 \\ 
   \phi_i^5  \\ 
  {\phi}^2 \phi_i^3 \\ 
  \phi_i  \sum_{j=1}^{N} \phi_j^4   \\ 
  \end{array} } \right) } &  \qquad  \gamma = \frac{1}{3N}
  \scriptsize \left( {\begin{array}{cccc}
  4(N+5) & 0 & 3 & 6 \\ 
  0 & 10(N-2) & 6(N-4) & 4(N-4) \\ 
  4(N-4) & 10 & 5(N+2) & 4 \\ 
  4(N-4) & 0 & 0 & 2(3N-2) 
  \end{array} } \right) \\ \normalsize 
\end{array} $

The dimensions of these representations add up to $\frac{N(N+1)(N+2)(N+3)(N+4)}{120}$, while the 5-index traceless symmetric $O(N)$ tensor decomposes under $H_N$ as
\small \begin{eqnarray}
  && {\small \Yvcentermath1  \yng(5)}=\frac{N(N+8)(N+2)(N+1)(N-1)}{120} =2 \times N\oplus 2\times N(N-2)\oplus\frac{N(N-1)(N-2)}{6} \nonumber \\
  &&\oplus {\frac{N(N-1)(N-4)}{2}}\oplus{\frac{N(N-1)(N-2)}{3}}\oplus{
  \left( {\begin{array}{cc}
  N  \\
  3 \\
  \end{array} } \right)(N-4)} \oplus
  {
  \left( {\begin{array}{cc}
  N  \\
  5 \\ 
  \end{array} } \right)} \ .
\end{eqnarray} \normalsize
They pass all the consistency checks in a way similar to Sec.\ref{n=3} for $N=1,2,4,\infty$. Namely, for $N=1$, it is easy to check that the four-by-four mixing matrix for the vectors has one zero eigenvalue while for $N=2, \infty$ it has one eigenvalue equal to 10/3 as required by Eq.\eqref{towerising}. We are not aware of any explicit results for $n\ge 5$ in the literature.

\subsection{$n \ge 6$} \label{feature}

Even though the techniques described in this paper allow us to compute the whole spectrum of $\gamma$'s up to arbitrary $n$, we stop giving exhaustive results and simply provide a qualitative description of the spectrum for $n \ge 6$.

At every $n$, all non-mixing scaling operators together with their $\gamma$'s, are given by Eqs. \eqref{eq5} and \eqref{gammas}. For instance, at $n=6$ these are\footnote{Again, all the indices of the operators listed here have to be understood being different, e.g. $i \neq j \neq...\neq m$.}
\begin{eqnarray} \label{six}
  & \phi_i \phi_j \phi_k \phi_l \phi_m \phi_n &\quad \gamma=\frac{10}{N} \nonumber \\
  & \phi_i \phi_j \phi_k \phi_l (\phi_m^2 - \phi_n^2) &\quad  \gamma=\frac{N+26}{3N} \nonumber \\
  & \phi_i \phi_j (\phi_k^2 - \phi_l^2)(\phi_m^2 - \phi_n^2) &\quad  \gamma=\frac{2(N+11)}{3N} \nonumber \\
  & (\phi_i^2 - \phi_j^2) (\phi_k^2 - \phi_l^2)(\phi_m^2 - \phi_n^2) &\quad  \gamma=\frac{N+6}{N} \nonumber \\
  & \phi_i \phi_j (\phi_k^3 \phi_l - \phi_l^3 \phi_k ) &\quad  \gamma=\frac{N+6}{N} \nonumber \\
  &  (\phi_i^2 - \phi_j^2) (\phi_k^3 \phi_l - \phi_l^3 \phi_k ) &\quad  \gamma=\frac{2(2N+7)}{3N} \nonumber \\
  & \phi_k^4 \phi_i^2+\phi_j^4 \phi_k^2+\phi_i^4 \phi_j^2-\phi_k^4 \phi_j^2-\phi_i^4 \phi_k^2-\phi_j^4 \phi_i^2 &\quad  \gamma=\frac{7N+2}{3N} 
\end{eqnarray}
The first six operators listed above correspond to bi-tableaux with a maximum of two rows, and thus their $\gamma$'s can be computed using Eq.\eqref{torre2}.
In fact, there they correspond to $\{p_1,p_2,p_3\}=\{6,0,0\},\{4,1,0\},\{2,2,0\},\{0,3,0\},\{2,0,1\},\{0,1,1\}$ respectively.

Moreover, at $n=6$ we find, for the first time, a unique operator whose $\gamma$ cannot be computed by using Eq.\eqref{torre2}. Therefore, in order to compute its $\gamma$, one has to resort to the full Eq.\eqref{gammas}. This is the last operator in Eq.\eqref{six}, which corresponds to the following bi-tableau with three rows in the left partition ($k=1, \ m_i=4, \ p_i=1$):
\ytableausetup
  {boxsize=1em, aligntableaux=center}
 \begin{equation}
\left( \  \begin{ytableau}
\empty &\empty &\none[...]& \empty \\
\empty& \none[] \\
\empty& \none[]  \\
 \end{ytableau}  {\huge{,}} 
 \ \emptyset
  \  \right) .
   \end{equation}

Returning to the general $n$, apart from non-mixing scaling operators, all the other irreducible representations will be given by those at the level $n-2$ and all the corresponding eigenvectors can be generated as explained in Sec.\ref{Irreducible representations of $H_N$}. At every even $n$, the lowest dimensional irreducible representations are given by the scalar sector. This is given by Eq.\eqref{scalarmix}, which implies that the number of scalars which mix at a given order $n$ is the number of partitions of $\frac{n}{2}$. 
Moreover, in this scalar sector at every level $n$ there is at least one $\gamma_n (N=1)=0$ needed in order to obtain the free Gaussian theory. Of course, this happens also in the case of the vectorial ($dim = N$) irreducible representations, which appear at every odd $n$.  
The scalar sector for $n=6$ is
\begin{equation} \label{scalar6}
  \left( \begin{array}{c}
  (\phi^2)\sum_{i} \phi_i^4 \\
  (\phi^2)^3\\
  \sum_{i} \phi_i^6  \\
  \end{array} \right) \ \ \ \gamma =
  \frac{1}{3N}\left( \begin{array}{ccc}
  2(5+4N) & 12(N-4) & 15 \\
  6 & 6(N+5) & 0 \\
  8(N-4) & 0 & 15(N-2)
  \end{array} \right)
\end{equation}
and it is easy to check that for $N=1$ one eigenvalue is $0$.

Further insights on the spectrum of anomalous dimensions can be gained by looking at
\begin{equation}
    W_n=\sum_{S_n}d_{S_n} \gamma_{S_n}
\end{equation}
where $d_{S_n}$ and $\gamma_{S_n}$ are the dimensions and the anomalous dimensions of the composite operators $S_n$, respectively, and the sum runs over all the irreducible representations at the level $n$.
In fact, from Eq.\eqref{eq7}, it follows that, for any $n$, $W_n$ has to be proportional to $N-1$. This can be easily checked for $n \le 5$ using our previous results, while for $n=6$, it can be checked by noting that the only irreducible representations which contribute to $W_6 (N=1)$ are Eq.\eqref{scalar6} and
\begin{equation}
 \left( \begin{array}{c}
  (\phi^2)(\phi_i^2-\phi_j^2)(\phi_k^2-\phi_l^2) \\
 (\phi_i^4-\phi_j^4)(\phi_k^2-\phi_l^2)+(\phi_i^2-\phi_j^2)(\phi_k^4-\phi_l^4)
  \end{array} \right) \ \ \ \gamma =
  \frac{1}{3N}\left( \begin{array}{cc}
  2(13+2N) & 12  \\
  4(N-4) & 2+7N
  \end{array} \right)
\end{equation}
which has dimension $2 \times \tfrac{1}{2}N(N-3)$.

Moreover, the values of $W_n$ exhibit an interesting pattern, \footnote{We thank H.Osborn for this observation.} 
\begin{align}
   &  W_2 = \frac{2}{3}(N-1),  \qquad  \qquad  \qquad \qquad \ \ W_3 = \frac{2}{3}(N-1)(N+2),  \\
   &    W_4 = \frac{1}{3}(N-1)(N+2)(N+3), \qquad  W_5 = \frac{1}{9}(N-1)(N+2)(N+3)(N+4) \nonumber
\end{align}
which is indicative of a general formula for $W_n$. 


We conclude this section by giving the decomposition of the 6-index traceless symmetric tensor of $O(N)$ under $H_N$:
\small \begin{eqnarray}
  && {\small \Yvcentermath1  \yng(6)}=\frac{N(N+10)(N+3)(N+2)(N+1)(N-1)}{6!} = 1\oplus 2\times (N-1)  
  \oplus 3\times \frac{N(N-1)}{2}   \nonumber \\ 
  && \oplus \frac{N(N-1)}{2} \oplus 2 \times \frac{N(N-1)(N-3)}{2}\oplus \frac{N(N-3)}{2}\oplus {\frac{N(N-1)(N-5)}{6}}\oplus \frac{N(N-1)(N-2)(N-5)}{4}  \nonumber \\
  && \oplus (N-5) 
  \left( {\begin{array}{cc}
  N  \\
  4 \\
  \end{array} } \right)
  \oplus 
  {
   \left( {\begin{array}{cc}
   N  \\
   6 \\
   \end{array} } \right)}
   \oplus 
  {
   \left( {\begin{array}{cc}  
   N  \\
   4 \\
  \end{array} } \right)}
  \oplus 
  3 {
  \left( {\begin{array}{cc}
   N  \\
   4 \\
  \end{array} } \right)}
  \oplus 
  (N-3) {
  \left( {\begin{array}{cc}
   N  \\
   2 \\
  \end{array} } \right)}
  \oplus 
  {\frac{(N-1)(N-2)}{2}} \ .
\end{eqnarray} \normalsize

\section{Towards complete CFT data} \label{cftdata}

The description of the spectrum of anomalous dimensions is an important step towards the computation of complete CFT data for the hypercubic model. The other important set of data is provided by the structure constants $C_{ijk}$ defined by
\begin{equation}
    \langle O_i(x) O_j(y) O_k(z) \rangle = \frac{C_{ijk}}{|x-y|^{\Delta_i+\Delta_j-\Delta_k}|y-z|^{\Delta_j+\Delta_k-\Delta_i}|z-x|^{\Delta_k+\Delta_i-\Delta_j}}
\end{equation}
The simplest class of $C_{ijk}$ corresponds to $C^{112}= \langle \phi_i \phi_j S_2\rangle$, where $S_2$ is a scaling operator of order $n=2$. These coefficients were already computed up to $\mathcal{O}(\epsilon^2)$ in \cite{Dey:2016mcs} using the conformal bootstrap, while all the $C^{113}$ are trivially $0$. We therefore present a novel computation for a sample of $C^{114}= \langle \phi_i \phi_j S_4\rangle$ coefficients.

For this purpose, techniques developed in \cite{Codello:2018nbe} and described in Sec.\ref{Computation} are again useful and allow the computation of several classes of leading order structure constants. Moreover, this task is noticeably simplified by the group-theoretic analysis of Sec.\ref{Irreducible representations of $H_N$}. Coefficients $C^{114}$ can be computed via the following formula \cite{Codello:2018nbe}:
\begin{equation}
    C^{114}= \frac{1}{8 \pi^2} V_{imnr} S_{jmnr}
\end{equation}
Here, $V_{imnr}$ and $S_{jmnr}$ are again given, respectively, by Eqs.\eqref{eq4} and \eqref{Stensor}.
We obtain
\begin{eqnarray} \label{structure}
 &  \phi_i \phi_j \phi_k \phi_l  & \qquad C^{114}=0 + \mathcal{O}(\epsilon^2) \qquad i \neq j \neq k \neq l \nonumber \\
 &  \phi_k^3 \phi_l-\phi_l^3 \phi_k  & \qquad C^{114}=\frac{(N-4)}{6N}(\delta_{kj}\delta_{li}-\delta_{lj}\delta_{ki})\epsilon+ \mathcal{O}(\epsilon^2) \qquad k \neq l \nonumber \\
  & (\phi^2)^2 -2 \sum_k \phi_k^4 & \qquad C^{114}=0+\mathcal{O}(\epsilon^2) \nonumber \\
  & (\phi^2)^2 +\frac{(N-4)}{3} \sum_k \phi_k^4 & \qquad C^{114}=\frac{2}{9N}(N^2+N-2)\delta_{ij} 
  \epsilon+\mathcal{O}(\epsilon^2)
\end{eqnarray}

Up to now, we have considered only scaling operators with no derivatives. Aside from diagrammatic techniques, the spectrum of $\gamma$'s for operators with derivatives can be computed using the conformal bootstrap as in \cite{Dey:2016mcs} or generalizing the techniques of Sec.\ref{Computation}. Work along this direction was initiated in \cite{Roumpedakis:2016qcg, Liendo:2017wsn} and we plan to investigate this generalization in the future. 

Another future direction is to extend this analysis to the next-to-leading order in the $\epsilon$-expansion. At the moment, this requires approaches quite different from those of Sec.\ref{Computation}, such as exploiting the conformal bootstrap machinery. It is an open problem to see if, by extending the EOM technique of Sec.\ref{Computation} to higher order correlation functions, one can access next-to-leading order in the $\epsilon$-expansion in a relatively simple way.

\section{Conclusion} \label{conclusion}

In this paper, we have computed the spectrum of anomalous dimensions at the leading order in $\epsilon$-expansion in the $O(N)$ model with cubic anisotropy in $d = 4 - \epsilon$. We derived the complete closed-form expression for the anomalous dimensions of the operators which do not undergo mixing effects and delineated the structure of the general solution to the mixing problem for arbitrary $n$. 
We have examined the general features of the spectrum uncovering the interesting pattern of the values for the sum of the anomalous dimensions weighted by the size of the corresponding representations pointing to the deeper structure.
A plethora of examples was provided, including the explicit solution for operators with $n\le 6$.

For this purpose, in Sec.\ref{Irreducible representations of $H_N$}, we have systematized, for the the first time in the literature, the representation theory for $H_N$ with the aim of providing a solid ground for future analyses of models with hypercubic symmetry. For instance, it can be useful for extending the conformal bootstrap analysis of model with $H_3$, recently performed in \cite{Kousvos:2018rhl, Stergiou:2018gjj}, to the general $H_N$ case. It would be also instrumental in extensions of our results to operators with derivatives and/or higher orders in $\epsilon$-expansion.


 As another future direction, it would be interesting to compute the spectrum of anomalous dimensions in models with different symmetries, e.g. in theories invariant under the hypertetrahedral symmetry group $H_{tetrahedral} \simeq \mathcal{S}_{N+1} \otimes \mathcal{Z}_2$ \cite{Osborn:2017ucf} or in the so-called \textit{Platonic field theories} recently studied in \cite{Zinati:2019gct}.

\section*{Acknowledgements}
It is our pleasure to thank Georgios Karagiannis, Alessio Maiezza, Blaženka Melić and Hugh Osborn for fruitful discussions and comments on the manuscript. This work was partially supported by the Croatian Science Foundation project number 4418 as well as European Union through the European Regional Development Fund - the Competitiveness and Cohesion Operational Programme (KK.01.1.1.06).

\end{document}